\documentclass[12pt]{article}
\usepackage{subfloat}
\pdfoutput = 1
\begin{document}
\date{Today}
\title{{\bf{\Large  Analytic study of properties of holographic superconductors in Born-Infeld electrodynamics}}}

\author{
{\bf {\normalsize Sunandan Gangopadhyay}$^{a,c,d}
$\thanks{sunandan.gangopadhyay@gmail.com, sunandan@bose.res.in}},
{\bf {\normalsize Dibakar Roychowdhury}
$^{b,}$\thanks{dibakar@bose.res.in}}\\
$^{a}$ {\normalsize Department of Physics, West Bengal State University, Barasat, India}\\
$^{b}$ {\normalsize  S.N. Bose National Centre for Basic Sciences,}\\{\normalsize JD Block, 
Sector III, Salt Lake, Kolkata 700098, India}\\[0.3cm]
$^{c}${\normalsize Visiting Associate in S.N. Bose National Centre for Basic Sciences,}\\
{\normalsize JD Block, Sector III, Salt Lake, Kolkata 700098, India}\\[0.3cm]
$^{d}${\normalsize Visiting Associate in Inter University Centre for Astronomy \& Astrophysics,}\\
{\normalsize Pune, India}\\[0.3cm]
}
\date{}

\maketitle

\begin{abstract}
{\noindent In this paper, based on the Sturm-Liouville eigenvalue problem, we analytically investigate several properties of holographic $s$-wave superconductors in the background of a Schwarzschild-AdS spacetime in the framework of 
Born-Infeld electrodynamics. Based on a perturbative approach, we explicitly find the relation between the critical temperature and the charge density and also the fact
that the Born-Infeld coupling parameter indeed affects the formation of scalar hair at low temperatures. Higher value of the Born-Infeld parameter results in a harder condensation to form. 
We further compute the critical exponent associated with the condensation near the critical temperature. The analytical results obtained 
are found to be in good agreement with the existing numerical results.  }
\end{abstract}
\vskip 1cm

\section{Introduction}
For the past sixty years, the BCS theory of superconductivity has been the most successful microscopic theory to describe weakly coupled superconductors with great accuracy \cite{bcs}. The basic principle that is responsible to exhibit superconductivity in these weakly coupled systems is the spontaneous breaking of $ U(1) $ symmetry at low temperatures. However, it has been realized for quite some time that there are some materials, like heavy fermion compounds or high $ T_c $ cuprates where the understanding of the pairing mechanism remains incomplete. The failure of BCS theory in order to understand such strongly coupled systems invites new theoretical inputs. One such input comes from the so called AdS/CFT correspondence.

The AdS/CFT duality has been a powerful tool to deal with strongly coupled systems. It provides an exact correspondence
between the gravity theory in a $ (d+1) $ dimensional AdS space time 
and a conformal field theory (CFT) residing on its $d$-dimensional boundary \cite{adscft1}-\cite{adscft4}. In recent years the AdS/CFT correspondence has been found to provide some meaningful theoretical insights in order to understand the physics of high $T_c$ superconductors. The holographic description of  $s$-wave superconductors basically consists of a black hole and a complex scalar field minimally coupled to an abelian gauge field. The formation of scalar hair below certain critical temperature ($T_c$) indicates the onset of a condensation in the dual CFTs. The mechanism that is responsible behind this condensation is the breaking of a local $U(1)$ symmetry near the event horizon of the black hole \cite{hs1}-\cite{hs5}.

Till date, most of the investigations that have been performed on various holographic superconductor models are based on the framework of usual Maxwell electrodynamics \cite{hs6}-\cite{hs18}.  
Recently, investigations have also been carried out in 
the framework of non-linear electrodynamics \cite{hs19}-\cite{hs22}. 
In particular, the effects of Born-Infeld electrodynamics (BI) on the holographic superconductors has been
studied  numerically in \cite{hs19}. The analysis is important in its own right as  
BI electrodynamics is one of the most important non-linear electromagnetic theory free from infinite self energies of charged point particles that arises in the Maxwell theory and it is also the only theory invariant under the electromagnetic duality transformations \cite{BI}. All these features are sufficiently motivating to study Einstein gravity in the frame work of BI electrodynamics \cite{plb2}-\cite{olivera}. In spite of several attempts, an analytic study of the properties of holographic superconductors in the framework of BI electrodynamics has however been missing in the existing literature.   

In this paper, based on the Sturm-Liouville (SL) eigenvalue problem, we attempt to provide an answer to some of the major issues related to holographic $s$-wave superconductors in the framework of BI electrodynamics. We analytically
show (upto first order in the BI coupling parameter) the relation between the critical temperature and the charge density and also the fact that
at low temperatures $(T<T_c)$, the condensation is indeed affected by the presence of BI coupling parameter. 
The critical exponent for the condensation near the critical temperature also 
comes out naturally in our analysis. Our results have been found to be in good agreement with the numerical results  existing in the literature\cite{hs19}. 
It is reassuring to note that all our calculations have been carried out in the probe limit.      

This paper is organized as follows. In section 2, we provide the basic holographic set up for the $s$-wave superconductors in the framework of BI electrodynamics, considering the background of a Schwarzschild-AdS spacetime.
In section 3, ignoring the back reaction of the dynamical matter field on the spacetime metric and using the perturbative technique, we compute the critical temperature in terms of a solution to the SL eigenvalue problem. In section 4,
we determine the temperature dependence of the condensate upto first order in the BI coupling parameter.  Finally, we conclude in section 5.

\section{Basic set up for $s$-wave superconductors}
Our construction of the holographic $s$-wave superconductor is based on the fixed
background of Schwarzschild-AdS spacetime.
The metric of a planar Schwarzschild-AdS black hole reads
\begin{eqnarray}
ds^2=-f(r)dt^2+\frac{1}{f(r)}dr^2+r^2(dx^2+dy^2)
\label{m1}
\end{eqnarray}
with
\begin{eqnarray}
f(r)=r^{2}-\frac{r_{+}^3}{r}
\label{metric}
\end{eqnarray}
in units in which the AdS radius is unity, i.e. $l=1$.
The Hawking temperature is related to the horizon radius ($r_+$) as
\begin{eqnarray}
T=\frac{3r_+}{4\pi}~.
\label{temp}
\end{eqnarray}
Let us now consider an electric field and a charged complex scalar
field in this fixed background. The corresponding
Lagrangian density can be expressed as
\begin{eqnarray}
\mathcal{L}=\mathcal{L}_{BI}-
|\nabla_{\mu}\psi-iqA_{\mu}\psi|^2-m^2|\psi|^2
\label{m10}
\end{eqnarray}
where $\psi$ is a charged complex scalar field, $\mathcal{L}_{BI}$
is the Lagrangian density of the Born-Infeld electrodynamics
\begin{eqnarray}
\mathcal{L}_{BI}=\frac{1}{b}\bigg(1-\sqrt{1+\frac{b F}{2}}\bigg).
\label{m11}
\end{eqnarray}
Here $F\equiv F_{\mu\nu}F^{\mu\nu}$ and $F_{\mu\nu}$ is the
non-linear electromagnetic tensor which satisfies the BI
equation
\begin{eqnarray}
\partial_{\mu}\bigg(\frac{\sqrt{-g}F^{\mu\nu}}{\sqrt{1+\frac{b F}{2}}}\bigg)=J^{\nu}
\label{m12}
\end{eqnarray}
with the BI coupling parameter $b$ indicating the difference between BI
and Maxwell electrodynamics. In the limit $b\rightarrow 0$, the Lagrangian
$\mathcal{L}_{BI}$ approaches to $-\frac{1}{4}F_{\mu\nu}F^{\mu\nu}$,
and one recovers the standard Einstein-Maxwell theory. 

\noindent In order to solve the equations of motion both for the complex scalar field and the electromagnetic field,
we adopt the following ansatz \cite{hs6}
\begin{eqnarray}
A_{\mu}=(\phi(r),0,0,0),\;\;\;\;\psi=\psi(r)
\label{vector}
\end{eqnarray}
which finally yields the equations of motion for the complex scalar
field $\psi(r)$ and electrical scalar potential $\phi(r)$ as
\begin{eqnarray}
\psi^{''}(r)+\left(\frac{f'}{f}+\frac{2}{r}\right)\psi'(r)
+\left(\frac{\phi^{2}(r)}{f^2}-\frac{m^2}{f}\right)\psi(r)=0\label{e1}
\end{eqnarray}
and
\begin{eqnarray}
\phi''(r)+\frac{2}{r}\phi'(r)\bigg(1-b\phi'^2 (r)\bigg)
-\frac{2\psi^2 (r)}{f}\phi(r)\bigg(1-b\phi'^2 (r)\bigg)^{3/2}=0\label{e2}
\end{eqnarray}
where prime denotes derivative with respect to $r$. In order to solve the
non-linear equations (\ref{e1}) and (\ref{e2}), we need
to seek the boundary condition for $\phi$ and $\psi$ near the black
hole horizon $r\sim r_+$ and at the spatial infinite
$r\rightarrow\infty$. The regularity condition at the horizon gives
the boundary conditions $\phi(r_+)=0$ and
$\psi=-\frac{3r_H}{2}\psi'$. 

\noindent Under the change of coordinates $z=r_{+}/r$,  the field equations become
\begin{eqnarray}
z\psi''(z)-\frac{2+z^3}{1-z^3}\psi'(z)
+\left[z\frac{\phi^{2}(z)}{r_{+}^{2}(1-z^3)^2}-\frac{m^2}{z(1-z^3)}\right]\psi(z)=0\label{e1a}
\end{eqnarray}
\begin{eqnarray}
\phi''(z)+\frac{2bz^3}{r_{+}^2 }\phi'^{3}(z)-\frac{2\psi^2 (z)}{z^2 (1-z^3)}
\left(1-\frac{bz^4}{r_{+}^2}\phi'^{2}(z)\right)^{3/2}\phi(z) =0\label{e1aa}
\end{eqnarray}
where prime now denotes derivative with respect to $z$. These equations are to be solved in the
interval $(0, 1)$, where $z=1$ is the horizon and $z=0$ is the boundary.
The boundary condition $\phi(r_+)=0$ now becomes $\phi(z=1)=0$.

\noindent Setting $m^2$ close to BF bound \cite{bf1}-\cite{bf2} , the asymptotic boundary conditions for
the scalar potential $\phi(z)$ and the scalar field $\psi(z)$ turn out to be
\begin{eqnarray}
\phi\approx\mu-\frac{\rho}{r}=\mu-\frac{\rho}{r_{+}}z
\label{b2}
\end{eqnarray}
\begin{eqnarray}
\psi\approx\frac{\psi^{(-)}}{r^{\Delta_{-}}}+\frac{\psi^{(+)}}{r^{\Delta_{+}}}\label{b1}
\end{eqnarray}
where 
\begin{eqnarray}
\Delta_{\pm}=\frac{3}{2}\pm\sqrt{\frac{9}{4}+m^2}
\label{dimension}
\end{eqnarray}
is the conformal dimension of the condensation operator $J$ in the boundary
field theory. 
The coefficients $\psi^{(-)}$ and $\psi^{(+)}$
correspond to the vacuum expectation values of the condensation
operator $J$ dual to the scalar field. Also $\mu$ and $\rho$ are 
interpreted as the chemical potential and the charge density of the dual theory
on the boundary.
Setting $m^2 =-2$ in eq.(\ref{dimension}), we have $\Delta_{-}=1$ and $\Delta_{+}=2$.
As in \cite{hs6}, we can impose the boundary condition that either $\psi^{(-)}$ or
$\psi^{(+)}$ vanish, so that the theory is stable in the asymptotic
AdS region.
In this paper, we shall set $\psi^{(+)}=0$ and $\langle J\rangle=\psi^{(-)}$. 

\section{Relation between critical temperature and charge density}
With the above set up in place, we are now in a position to investigate the relation between the critical temperature and the charge density. 

\noindent At the critical temperature $T_c$, $\psi=0$, so the field equation (\ref{e1aa})
for the electrostatic potential $\phi$ reduces to
\begin{eqnarray}
\phi''(z)+\frac{2bz^3}{r_{+(c)}^2}\phi'^{3}(z)=0.
\label{e1b}
\end{eqnarray}
To solve the above equation, we set $\phi'(z)=\xi(z)$ and obtain 
\begin{eqnarray}
\xi'(z)+\frac{2bz^3}{r_{+(c)}^2 }\xi^{3}(z)=0.
\label{eqn}
\end{eqnarray}
Integrating the above equation in the interval $[0, 1]$, we get 
\begin{eqnarray}
\frac{1}{\xi^{2}(1)}-\frac{1}{\xi^{2}(0)}=\frac{b}{r_{+(c)}^2}
\label{eqn1}
\end{eqnarray}
where we have used the fact that $\xi=\xi(0)$ at $z=0$ and $\xi=\xi(1)$ at $z=1$. 

\noindent At $z=0$, from eq.(\ref{b2}) we have
\begin{eqnarray}
\phi'|_{z=0}=\xi(0) &\approx&-\frac{\rho}{r_{+}}
=-\frac{\rho}{r_{+(c)}} \quad at ~ T=T_c ~.
\label{b22}
\end{eqnarray}
From eq(s)(\ref{eqn1}, \ref{b22}), we obtain
\begin{eqnarray}
\frac{1}{\xi^{2}(1)}=\frac{b}{r_{+(c)}^2}+ \left(\frac{r_{+(c)}}{\rho}\right)^2 ~.
\label{eqn1a}
\end{eqnarray}
Hence, integrating eq.(\ref{eqn}) in the interval $[1, z]$ and using eq.(\ref{eqn1a}) leads to
\begin{eqnarray}
\xi(z)=\phi'(z)=-\frac{\lambda r_{+(c)}}{\sqrt{1+b\lambda^2 z^4}}
\label{b20}
\end{eqnarray}
where 
\begin{eqnarray}
\lambda=\frac{\rho}{r_{+(c)}^2}
\label{lam}
\end{eqnarray}
and the negative sign has been taken before the square root in the expression for $\phi'(z)$ since $\phi'(0)$
is negative at $z=0$ (eq.\ref{b22}).

\noindent Integrating eq.(\ref{b20}) again from $z'=1$ to $z'=z$, we obtain
\begin{eqnarray}
\phi(z)=\int_{1}^{z}\frac{\lambda r_{+(c)}}{\sqrt{1+b\lambda^2 z'^{4}}}dz'
\label{lam1}
\end{eqnarray}
where we have used the fact that $\phi(z=1)=0$.

\noindent The above integral is not doable exactly and hence we shall expand the integrand binomially
and keep terms upto $\mathcal{O}(b)$ to get
\begin{eqnarray}
\phi(z)=\lambda r_{+(c)}(1-z)\left\{1-\frac{b\lambda^2}{10}(1+z+z^2 +z^3 +z^4)\right\}\quad,\quad b\lambda^2 < 1.
\label{sol}
\end{eqnarray}
Note that the above solution satisfies eq.(\ref{e1b}) upto $\mathcal{O}(b)$ along with the boundary condition $\phi(z=1)=0$.

\noindent Using the above solution, we find that as $T\rightarrow T_c$, 
the field equation for the scalar field $\psi$ approaches the limit 
\begin{eqnarray}
-\psi''(z)+\frac{2+z^3}{z(1-z^3)}\psi'(z)
+\frac{m^2}{z^2 (1-z^3)}\psi(z)=\frac{\lambda^2 }{(1+z+z^2)^2}\left\{1-\frac{b\lambda^2}{5}\zeta(z)\right\}\psi(z)
\label{e001}
\end{eqnarray}
where $\zeta(z)=(1+z+z^2 +z^3 +z^4)$.

\noindent Near the boundary, we define \cite{hs7}
\begin{eqnarray}
\psi(z)=\frac{\langle J\rangle}{\sqrt 2 r_+} zF(z)
\label{sol1}
\end{eqnarray}
where $F(0)=1$.
Substituting this form of $\psi(z)$ in eq.(\ref{e001}), we obtain
\begin{eqnarray}
- F''(z) + \frac{3z^2}{1-z^3}F'(z) + \frac{z}{1-z^3}F(z)
&=&\frac{\lambda^2 }{(1+z+z^2)^2}\left\{1-\frac{b\lambda^2}{5}\zeta(z)\right\}F(z) \nonumber\\
&\approx&\frac{\lambda^2 }{(1+z+z^2)^2}\left\{1-\frac{b(\lambda^2|_{b=0})}{5}\zeta(z)\right\}F(z)\nonumber\\
\label{eq5b}
\end{eqnarray}
to be solved subject to the boundary condition $F' (0)=0$. Note that in the second line we have used the fact
that $b\lambda^2 =b(\lambda^2|_{b=0}) +\mathcal{O}(b^2)$, where $\lambda^2|_{b=0}$ is the value of $\lambda^2$ for $b=0$.

\noindent The above equation can be put in the Sturm-Liouville form (see eq.(\ref{app6}) in the appendix)
with 
\begin{eqnarray}
p(z)=1-z^3~,~ q(z)=z~,~r(z)=\frac{1-z}{1+z+z^2}\left\{1-\frac{b(\lambda^2|_{b=0})}{5}\zeta(z)\right\}. 
\label{i1}
\end{eqnarray}
With the above identification, we now write down the eigenvalue $\lambda^2$ which minimizes the expression (see appendix)
\begin{eqnarray}
\lambda^2 = \frac{\int_0^1 dz\ \{ (1-z^3) [F'(z)]^2 + z [F(z)]^2 \} }{\int_0^1 dz \ \frac{1-z}{1+z+z^2}
\left\{1-\frac{b(\lambda^2|_{b=0})}{5}\zeta(z)\right\} [F(z)]^2}~.
 \label{eq5abc}
\end{eqnarray}
To estimate it, we use the following trial function
\begin{eqnarray}
F= F_\alpha (z) \equiv 1 - \alpha z^2
\label{eq50}
\end{eqnarray}
which satisfies the conditions $F(0)=1$ and $F'(0)=0$.

\noindent For $b=0$, we obtain
\begin{eqnarray}
\lambda_\alpha^2|_{b=0} = \frac{6-6\alpha + 10\alpha^2}{2\sqrt 3 \pi - 6\ln 3
+ 4 (\sqrt 3 \pi +3\ln 3 - 9)\alpha + (12\ln 3 - 13)\alpha^2} 
\end{eqnarray}
which attains its minimum at $\alpha \approx 0.2389$. Hence, we have
\begin{eqnarray}
\lambda^2|_{b=0} \approx \lambda_{0.2389}^2|_{b=0} \approx 1.268 
\end{eqnarray}
to be compared with the exact value $\lambda^2|_{b=0} = 1.245$.
The critical temperature therefore reads 
\begin{eqnarray}
T_c = \frac{3}{4\pi} r_{+(c)} = \frac{3}{4\pi} \sqrt{\frac{\rho}{\lambda|_{b=0}}}\approx 0.225\sqrt\rho 
\label{eqTc}
\end{eqnarray}
which is in very good agreement with the exact $T_c = 0.226\sqrt\rho$ \cite{hs6}.

\noindent For $b=0.1$, we obtain
\begin{eqnarray}
\lambda_\alpha^2|_{b=0.1} = \frac{\frac{1}{2}-\frac{\alpha}{2}+\frac{5\alpha^2}{6}}
{0.344-0.082\alpha +0.014\alpha^2} 
\end{eqnarray}
which attains its minimum at $\alpha \approx 0.2402$. Hence, we have
\begin{eqnarray}
\lambda^2|_{b=0.1} \approx \lambda_{0.2402}^2|_{b=0.1} \approx 1.317. 
\end{eqnarray}
The critical temperature therefore reads 
\begin{eqnarray}
T_c = \frac{3}{4\pi} r_{+(c)} = \frac{3}{4\pi} \sqrt{\frac{\rho}{\lambda|_{b=0.1}}}\approx 0.223\sqrt\rho 
\label{eqTc}
\end{eqnarray}
which is in very good agreement with the exact $T_c = 0.224\sqrt\rho$ \cite{hs19}.

\noindent For $b=0.2$, we obtain
\begin{eqnarray}
\lambda_\alpha^2|_{b=0.2} = \frac{\frac{1}{2}-\frac{\alpha}{2}+\frac{5\alpha^2}{6}}
{0.330-0.077\alpha +0.013\alpha^2} 
\end{eqnarray}
which attains its minimum at $\alpha \approx 0.2417$. Hence, we have
\begin{eqnarray}
\lambda^2|_{b=0.2} \approx \lambda_{0.2417}^2|_{b=0.2} \approx 1.37. 
\end{eqnarray}
The critical temperature therefore reads 
\begin{eqnarray}
T_c = \frac{3}{4\pi} r_{+(c)} = \frac{3}{4\pi} \sqrt{\frac{\rho}{\lambda|_{b=0.2}}}\approx 0.221\sqrt\rho 
\label{eqTc}
\end{eqnarray}
which is once again in very good agreement with the exact $T_c = 0.222\sqrt\rho$ \cite{hs19}.

\noindent For $b=0.3$, we obtain
\begin{eqnarray}
\lambda_\alpha^2|_{b=0.3} = \frac{\frac{1}{2}-\frac{\alpha}{2}+\frac{5\alpha^2}{6}}
{0.317-0.072\alpha +0.012\alpha^2} 
\end{eqnarray}
which attains its minimum at $\alpha \approx 0.2432$. Hence, we have
\begin{eqnarray}
\lambda^2|_{b=0.3} \approx \lambda_{0.2432}^2|_{b=0.3} \approx 1.43. 
\end{eqnarray}
The critical temperature therefore reads 
\begin{eqnarray}
T_c = \frac{3}{4\pi} r_{+(c)} = \frac{3}{4\pi} \sqrt{\frac{\rho}{\lambda|_{b=0.3}}}\approx 0.218\sqrt\rho 
\label{eqTc}
\end{eqnarray}
which is also in very good agreement with the exact $T_c = 0.219\sqrt\rho$ \cite{hs19}.

\section{Critical exponent and condensation values}
In this section, we shall compute the condensation values of the condensation operator $J$ in the boundary field theory.

\noindent Away from (but close to) the critical temperature $T_c$, the field equation (\ref{e1aa}) for $\Phi$ becomes (using eq.(\ref{sol1}))
\begin{eqnarray}
\phi''(z) +\frac{2bz^3}{r_{+}^2}\phi'^{3}(z)&=&\frac{\langle J\rangle^2}{r_+^{2}}\mathcal{B}(z)\phi(z)\label{aw1} \\
\mathcal{B}(z)&=&\frac{F^{2}(z)}{1-z^3}\left(1-\frac{3bz^4}{2r_{+}^2}\phi'^{2}(z)\right)+\mathcal{O}(b^2)\nonumber
\end{eqnarray}
where the parameter $\langle J\rangle^2/r_+^{2}$ is small.
We may now expand $\phi(z)$ in the small parameter $\langle J\rangle^2/r_+^{2}$ as
\begin{eqnarray}
\frac{\Phi}{r_+}=\lambda (1-z)\left\{1-\frac{b\lambda^2}{5}\zeta(z)\right\}+ \frac{\langle J\rangle^2}{r_+^{2}} \chi(z) 
+\dots
\label{aw2} 
\end{eqnarray}
From eq(s)(\ref{aw1}, \ref{aw2}) (keeping terms upto $\mathcal{O}(b)$), we obtain the equation for
the correction $\chi(z)$ near the critical temperature
\begin{eqnarray}
\chi''(z) +6b\lambda^2 z^3 \chi'(z) = \lambda \frac{F^{2}(z)}{1+z+z^2}\left\{1-\frac{b\lambda^2}{10}(\zeta(z)+15z^4)\right\}
\label{aw3}
\end{eqnarray}
with $\chi(1)=0=\chi'(1)$. Multiplying this equation by $e^{3b\lambda^2 z^{4}/2}$, we get
\begin{eqnarray}
\frac{d}{dz}\left(e^{3b\lambda^2 z^{4}/2}\chi'(z)\right) = \lambda e^{3b\lambda^2 z^{4}/2}
\frac{F^{2}(z)}{1+z+z^2}\left\{1-\frac{b\lambda^2}{10}(\zeta(z)+15z^4)\right\}~.
\label{aw4}
\end{eqnarray}
Integrating both sides of the above equation between $z=0$ to $z=1$, we obtain
\begin{eqnarray}
\chi'(0)=-\lambda\int_{0}^{1}dz~e^{3b\lambda^2 z^{4}/2}
\frac{F^{2}(z)}{1+z+z^2}\left\{1-\frac{b\lambda^2}{10}(\zeta(z)+15z^4)\right\}~.
\label{aw5}
\end{eqnarray}
Now from eq(s)(\ref{b2}, \ref{aw2}), we have
\begin{eqnarray}
\frac{\mu}{r_+}-\frac{\rho}{r_{+}^2}z&=&\lambda (1-z)\left\{1-\frac{b\lambda^2}{5}\zeta(z)\right\}+
\frac{\langle J\rangle^2}{r_+^{2}}\chi(z) \nonumber\\
&=&\lambda (1-z)\left\{1-\frac{b\lambda^2}{5}\zeta(z)\right\}+
\frac{\langle J\rangle^2}{r_+^{2}}(\chi(0)+z\chi'(0)+\dots)
\label{aw6}
\end{eqnarray}
where in the second line we have expanded $\chi(z)$ about $z=0$. Comparing the coefficient of $z$ on both sides
of the equation, we obtain
\begin{eqnarray}
\frac{\rho}{r_{+}^2}=\lambda-\frac{\langle J\rangle^2}{r_+^{2}}\chi'(0).
\label{aw7}
\end{eqnarray}
Substituting $\chi'(0)$ from eq.(\ref{aw5}) in the above equation, we get
\begin{eqnarray}
\frac{\rho}{r_{+}^2}=\lambda\left\{1+\frac{\langle J\rangle^2}{r_+^{2}}\mathcal{A}\right\}.
\label{aw8}
\end{eqnarray}
where
\begin{eqnarray}
\mathcal{A}=\int_{0}^{1}dz~e^{3b\lambda^2 z^{4}/2}
\frac{F^{2}(z)}{1+z+z^2}\left(1-\frac{b\lambda^2}{10}(\zeta(z)+15z^4)\right).
\label{aw9}
\end{eqnarray}
Finally using eq(s)(\ref{temp}, \ref{lam}), we get the following expression for $\langle J\rangle$
\begin{eqnarray}
\langle J\rangle=\gamma T_{c}\sqrt{1-\frac{T}{T_c}}
\label{aw10}
\end{eqnarray}
where
\begin{eqnarray}
\gamma=\frac{4\pi\sqrt{2}}{3\sqrt{\mathcal{A}}}~.
\label{aw11}
\end{eqnarray}
For $b=0$, computing $\mathcal{A}$ with $\alpha=0.2389$, we get $\gamma\approx8.07$
which is in good agreement with the exact result $\gamma\approx9.31$ \cite{hs6}.
For $b=0.1$, computing $\mathcal{A}$ with $\alpha=0.2402$, we get $\gamma\approx8.19$
which is in good agreement with the exact result $\gamma\approx9.48$ \cite{hs19}.
For $b=0.2$, computing $\mathcal{A}$ with $\alpha=0.2417$, we get $\gamma\approx8.33$
which is in good agreement with the exact result $\gamma\approx9.62$ \cite{hs19}.
For $b=0.3$, computing $\mathcal{A}$ with $\alpha=0.2432$, we get $\gamma\approx8.54$
which is in good agreement with the exact result $\gamma\approx9.74$ \cite{hs19}.

Let us summarize the results obtained in sections 3 and 4 in the table below:
\begin{table}[htb]
\caption{A comparison of the analytical and numerical results for the critical temperature and the expectation value of the condensation operator}   
\centering                          
\begin{tabular}{c c c c c c c}            
\hline\hline                        
$b$ & $\zeta_{SL}(=\frac{3}{4\pi}\sqrt{\frac{1}{\lambda_{min}}})$  & $\zeta_{Numerical}$ &$\gamma_{SL}(=\frac{4\pi\sqrt{2}}{3\sqrt{\mathcal{A}}})$ & $\gamma_{Numerical} $ &  \\ [0.05ex]
\hline
0 & 0.225 & 0.226 & 8.07 & 9.31 \\
0.1 & 0.223 & 0.224 & 8.19 & 9.48 \\                              
0.2 & 0.221 & 0.222 & 8.33 & 9.62 \\
0.3 & 0.218 &0.219 & 8.54 & 9.74 \\ [0.05ex]  
\hline                              
\end{tabular}\label{E1}  
\end{table}


\section{Conclusions}
In this paper, based on the Sturm-Liouville eigenvalue problem, we perform analytic computation of holographic $ s $-wave superconductors in Born-Infeld electrodynamics upto first order in the Born-Infeld coupling parameter. The relation between the critical temperature and the charge density has been obtained through an iterative procedure. It is further observed that the Born-Infeld coupling parameter decreases the critical temperature of the condensate indicating that it is harder for the scalar condensation to form in Born-Infeld electrodynamics. Our results are in very good agreement with the existing numerical results \cite{hs19}. The critical exponent of the condensation also comes out to be $ 1/2 $  which is the universal value in the mean field theory.

\section*{Acknowledgments} DR would like to thank CSIR for financial support.

\section*{Appendix: Sturm-Liouville problem and calculus of variations}

Let us consider the determination of stationary values of the quantity $\lambda$
defined by the ratio
\begin{eqnarray}
\lambda=\frac{\int_{a}^{b}\left\{p(x)(y'(x))^2 -q(x)y^2 (x)\right\}dx}{\int_{a}^{b} r(x)y^2 (x) dx}\equiv
\frac{I_1}{I_2}
\label{app1}
\end{eqnarray}
where $p(x)$, $q(x)$ and $r(x)$ are known functions of $x$ and prime denotes derivative with respect to $x$.

\noindent Varying $\lambda$ with respect to $y(x)$, we get
\begin{eqnarray}
\delta\lambda=\frac{1}{I_2}(\delta I_1 -\lambda\delta I_2)~.
\label{app2}
\end{eqnarray}
Computation of $\delta I_1$ and $\delta I_2$ yield
\begin{eqnarray}
\delta I_1=2\left\{p(x)y'(x)\delta y(x)\right\}|_{a}^{b} - 2\int_{a}^{b}\left\{(p(x)y'(x))' +q(x)y(x)\right\}\delta y(x) dx
\label{app3}
\end{eqnarray}
\begin{eqnarray}
\delta I_2=2\int_{a}^{b}r(x)y(x)\delta y(x) dx~.
\label{app4}
\end{eqnarray}
Substituting eq(s)(\ref{app3}, \ref{app4}) in eq.(\ref{app2}), we get
\begin{eqnarray}
\delta\lambda=\frac{2}{I_2}\left\{(p(x)y'(x)\delta y(x))|_{a}^{b} -\int_{a}^{b}\left[(p(x)y'(x))' +q(x)y(x)+\lambda r(x) y(x)
\right]\delta y(x) dx\right\}~.
\label{app5}
\end{eqnarray}
Hence, the condition $\delta\lambda=0$ for the stationary values of $\lambda$ leads to the Euler equation
\begin{eqnarray}
\frac{d}{dx}\left(p(x)\frac{dy(x)}{dx}\right)+q(x)y(x)+\lambda r(x) y(x)=0
\label{app6}
\end{eqnarray}
and to the following boundary conditions:
\begin{eqnarray}
(p(x)y'(x))|_{x=a}=0 \quad or \quad y(a)~ prescribed\nonumber\\
(p(x)y'(x))|_{x=b}=0 \quad or \quad y(b)~ prescribed.
\label{app7}
\end{eqnarray}


\end{document}